\documentclass[useAMS,usenatbib]{mn2e}
\usepackage{natbib}
\usepackage{amsmath,amsfonts}
\usepackage{epsfig}
\usepackage{mathptm}

\newcommand{\vect}[1]{{\mathbfit{#1}}}

\def\reff@jnl#1{{\rm#1\/}}

\def\aj{\reff@jnl{AJ}}                  
\def\araa{\reff@jnl{ARA\&A}}            
\def\apj{\reff@jnl{ApJ}}                        
\def\apjl{\reff@jnl{ApJ}}               
\def\apjs{\reff@jnl{ApJS}}              
\def\ao{\reff@jnl{Appl.Optics}}         
\def\apss{\reff@jnl{Ap\&SS}}            
\def\aap{\reff@jnl{A\&A}}               
\def\aapr{\reff@jnl{A\&A~Rev.}}         
\def\aaps{\reff@jnl{A\&AS}}             
\def\azh{\reff@jnl{AZh}}                        
\def\baas{\reff@jnl{BAAS}}              
\def\jrasc{\reff@jnl{JRASC}}            
\def\memras{\reff@jnl{MmRAS}}           
\def\mnras{\reff@jnl{MNRAS}}            
\def\pra{\reff@jnl{Phys.Rev.A}}         
\def\prb{\reff@jnl{Phys.Rev.B}}         
\def\prc{\reff@jnl{Phys.Rev.C}}         
\def\prd{\reff@jnl{Phys.Rev.D}}         
\def\prl{\reff@jnl{Phys.Rev.Lett}}      
\def\pasp{\reff@jnl{PASP}}              
\def\pasj{\reff@jnl{PASJ}}              
\def\qjras{\reff@jnl{QJRAS}}            
\def\skytel{\reff@jnl{S\&T}}            
\def\solphys{\reff@jnl{Solar~Phys.}}    
\def\sovast{\reff@jnl{Soviet~Ast.}}     
\def\ssr{\reff@jnl{Space~Sci.Rev.}}     
\def\zap{\reff@jnl{ZAp}}                        
\def\nat{\reff@jnl{Nature}}             

\label{firstpage}
\title[Cosmological parameter estimation and Bayesian model comparison]
{ Cosmological parameter estimation and Bayesian model comparison using VSA data}

\author[An\v{z}e Slosar et al.] {An\v{z}e Slosar$^1$, Pedro Carreira$^2$, 
  Kieran Cleary$^2$, Rod D. Davies$^2$, Richard J. Davis$^2$,
   \newauthor Clive Dickinson$^2$,   Ricardo Genova-Santos$^3$, Keith Grainge$^1$, Carlos M. Guti\'{e}rrez$^3$, 
 \newauthor Yaser A. Hafez$^2$, Michael
  P. Hobson$^1$, Michael E.  Jones$^1$, R\"udiger
  Kneissl$^1$,  \newauthor Katy Lancaster$^1$, Anthony Lasenby$^1$, J. P.
  Leahy$^2$, Klaus Maisinger$^1$,  Phil J. Marshall$^1$,  \newauthor Guy
  G. Pooley$^1$, Rafael Rebolo$^{3,4}$, Jos\'e Alberto
  Rubi\~no-Martin$^{3,\ddagger}$, Ben Rusholme$^{1,\star}$,  \newauthor Richard
  D. E. Saunders$^1$,  Richard Savage$^1$, Paul F. Scott$^1$,
  Pedro J. Sosa Molina$^3$,  \newauthor Angela C. Taylor$^1$, David
  Titterington$^1$, Elizabeth Waldram$^1$,   Robert A.
  Watson$^{2,\dagger}$  \newauthor and Althea Wilkinson$^2$. 
  \\
$^1$  Astrophysics Group, Cavendish Laboratory, Madingley Road,
Cambridge CB3 0HE, UK. \\
$^2$ Jodrell Bank Observatory, Macclesfield, Cheshire SK11 9DL, UK. \\
  $^3$Instituto de Astrof{\'i}sica de Canarias, 38200 La Laguna,
  Tenerife, Spain.\\ 
$^4$Consejo Superior de Investigaciones
  Cient{\'{\i}}ficas, Spain \\ 
$^{\dagger}$Present address: Instituto
  de Astrof{\'{\i}}sica de Canarias.\\
$^{\star}$Present address:
  Stanford University, Palo Alto, CA, USA\\
$^{\ddagger}$Present address: Max-Planck Institut f\"ur Astrophysik, Garching,
Germany}

\date{Accepted ---; received ---; in original form \today}
\pagerange{\pageref{firstpage}--\pageref{lastpage}}
\pubyear{2002}

\begin{document}
\maketitle

\begin{abstract}
We constrain the basic comological parameters using the
first observations by the Very Small Array (VSA) in its extended
configuration, together with existing cosmic microwave background 
data and other cosmological observations. 
We estimate cosmological parameters for
four different models of increasing complexity.
In each case, careful consideration is given to implied priors and the
Bayesian evidence is calculated in order to perform model
selection. We find that the data are most convincingly explained by a
simple flat $\Lambda\rm{CDM}$ cosmology without tensor modes. In this
case, combining just the VSA and COBE data sets yields the 68 per cent
confidence intervals $\Omega_{\rm{b}}h^2=0.034 ^{+0.007} _{-0.007}$,
$\Omega_{\rm{dm}}h^2=0.18 ^{+0.06}_{-0.04}$, $h=0.72^{+0.15}_{-0.13}
$, $n_s=1.07 ^{+0.06 }_{-0.06}$ and $\sigma_8=1.17 ^{ +0.25 }_{
-0.20}$.  The most general model considered includes spatial curvature,
tensor modes, massive neutrinos and a parameterised equation of state
for the dark energy. In this case, by combining all recent
cosmological data, we find, in particular, 95 percent limit on the
tensor-to-scalar ratio $R < 0.63$ and on the fraction of massive
neutrinos $f_\nu < 0.11$; we also obtain the 68 per cent confidence
interval $w=-1.06^{+0.20}_{-0.25}$ on the equation of state of dark
energy.
\end{abstract}


\section{Introduction}

In the past two years, a number of experiments have produced 
accurate measurements of the power spectrum of anisotropies in the 
cosmic microwave background (CMB) radiation on a range of angular
scales (Hanany et al. 2002; Netterfield et al. 2002; Halverson et
al. 2002; Sievers et al. 2002; Benoit et al. 2002).
These data, together with other cosmological observations, 
have been used to place increasingly tight constraints on the values of 
cosmological parameters in current models of the formation and
evolution of structure in the Universe.

In this letter, we repeat this process with the inclusion of
the latest observations from the Very Small Array (VSA) in its extended
configuration. Results from the VSA in its compact configuration have 
already been presented in Watson et al. (2002), Taylor et
al. (2002), Scott et al. (2002) and Rubi\~no-Martin et al. (2002) (hereafter Papers I - IV). 
In Grainge et al. (2002, hereafter Paper V) these data are combined
with the new extended configuration observations to
produce a combined power spectrum with 16 spectral bins spanning 
the range $\ell=160-1400$.  This joint set of observed 
band-powers provides powerful new constraints on cosmological
parameters. In this letter we extend the traditional
likelihood approach of previous analyses to a fully Bayesian 
treatment, including careful consideration of our
knowledge of cosmological parameters prior to the inclusion of any data, 
and the calculation of Bayesian evidences to perform model comparisons. 

\begin{table}
\begin{center}
\caption{The priors assumed for the basic 
parameters common to all four cosmological models under consideration.
The notation $(a,b)$ for parameter $x$ denotes a top-hat prior in the
range $a<x<b$}
\label{prior}
\begin{tabular}{lc}
\hline\hline 
Basic parameter & Prior \\
\hline
$\omega_b$ & $(0.005,0.80)$ \\
$\omega_{dm}$ & $(0.01,0.9)$ \\
$h$ &  $(0.4,1.0)$ \\
$n_s$ &  $(0.5,1.5)$ \\
$z_{re}$ & $(4,20)$\\
$10^{10} A_s$ & $(0,100)$ \\
\hline\hline
\end{tabular}
\end{center}
\end{table}

\section{Models, methods and priors}\label{models}

We restrict our attention to cosmological models in which the initial
density fluctuations are adiabatic with a simple power-law spectrum;
such perturbations are naturally produced in the standard single-field
inflationary model.  We assume the contents of the Universe to consist
of three components: baryonic matter, dark matter and vacuum energy,
with (present day) densities denoted by $\Omega_{\rm b}$, $\Omega_{\rm
dm}$ and $\Omega_\Lambda$ respectively, measured as a fraction of the
critical density required to make the Universe spatially-flat (with
$\Omega_{\rm b} + \Omega_{\rm dm} + \Omega_\Lambda=\Omega_{\rm tot}$).

\subsection{Model parameterisation and priors}

The parameterisation of the cosmological model can be performed in
numerous ways, although it is generally preferable to 
use physically-motivated parameters along principal degeneracy 
directions. To this end, in the most general case, we describe the
cosmological model using the following 11 parameters:
the Hubble parameter $h$ (defined as $H_0 = h 
\times 100\,{\rm km\,s^{-1}\,Mpc^{-1}}$); the physical baryon density
$\omega_{\rm b} \equiv \Omega_{\rm b}h^2$; the physical dark matter density
$\omega_{\rm dm} \equiv \Omega_{\rm dm}h^2$; the curvature density
$\Omega_{\rm k}=1-\Omega_{\rm tot}$; the fraction $f_\nu$ of dark
matter in the form of massive neutrinos; the parameter $w$ 
describing the equation of state of the dark energy ($p = w\rho$),
the redshift of (instantaneous) reionisation $z_{\rm re}$;
the amplitude of scalar modes $A_s$; the spectral index of
scalar modes $n_s$; the amplitude ratio $R$ of tensor to scalar modes 
and the spectral index of the tensor modes $n_t$.  

We consider four separate models of increasing 
complexity, in which a successively larger number of the above 
parameters are are allowed to vary. Model A assumes
spatial flatness, no massive neutrinos and no tensor modes;
Model B includes the possibility of non-zero curvature; Model C
additionally allows for the presence of tensor modes; and finally Model D
also allows for a contribution to the dark matter in the form
of massive neutrinos and also a variable $w$. The variable parameters common
to all four models are listed in Table~\ref{prior}, together with the
top-hat priors assumed for each. The values and priors assumed for
the other parameters are shown in Table~\ref{prior2} for each of the
four models under consideration.

\begin{table}
\begin{center}
\caption{The values and priors of assumed for the 
basic parameters defining the four cosmological models described in the
text. The notation $(a,b)$ for parameter $x$ denotes a top-hat prior in the
range $a<x<b$.}
\label{prior2}
\begin{tabular}{lcccc}
\hline\hline
& Model A & Model B & Model C & Model D    \\
\hline
$\Omega_k$ &$0$ & $(-0.25, 0.25)$ &  $(-0.25, 0.25)$ &  $(-0.25, 0.25)$ \\
$f_\nu$  & $ 0 $ & $ 0 $   & $ 0 $ & $ (0,0.2)$ \\
$w$ &  $ -1 $ & $ -1 $ & $ -1 $ & $ (-1.5,0) $ \\
$R$ & $  0$ & $  0 $ & $ (0,2) $ &  $ (0,2) $ \\
$n_t$ &        &         &    $(-1,0)$ & $(-1,0)$  \\
\hline\hline
\end{tabular}
\end{center}
\end{table}

The basic parameters described above completely define the 
models considered. It is of interest, however, also to
consider the following derived parameters: $\Omega_{\rm b}$,
$\Omega_{\rm dm}$, $\Omega_{\rm m}=\Omega_{\rm b}+\Omega_{\rm dm}$, 
$\Omega_{\Lambda}$, the age of the universe, the present day rms fluctuation 
in 8 $h^{-1}\, \rm Mpc$ spheres as predicted by linear
theory $\sigma_8$, and the optical depth to the surface of the last
scattering $\tau$. In addition to the priors listed in
Table~\ref{prior} and \ref{prior2}, we also impose the additional
constraints $\Omega_\Lambda \geq 0$ and a top-hat prior on the age of the
Universe lying between 10 and 20 Gyr.

\subsection{Bayesian analysis using MCMC samping}

Our approach to Bayesian parameter estimation and model selection
makes use of Markov-Chain Monte Carlo (MCMC) sampling to explore the
posterior distribution of the cosmological parameters (see e.g. Paper
IV). For any given model $M$, samples are drawn from the
(unnormalised) posterior distribution given by the product of the
likelihood and prior, i.e.  $\Pr(\vect{d}|\btheta,M)\Pr(\btheta|M)$,
where $\vect{d}$ denotes the data under analysis and $\btheta$ denotes
the parameters defining the model. The likelihood function is
evaluated in the same manner as discussed in Paper IV, and the joint
prior is simply the product of the individual priors discussed in the
previous section.

The particular implementation of the MCMC approach used is a slightly
adapted version of the \texttt{Cosmo-mc} software package (Lewis \&
Bridle 2002).  This sampler uses \texttt{CAMB} (Lewis, Challinor \&
Lasenby 2000) as its underlying power spectrum engine, and is specifically
tailored to the analysis of CMB data.  It achieves high acceptance
rates and rapid convergence by using a proposal function that exploits
the difference between `fast' and `slow' parameters in
\texttt{CAMB}.

\begin{figure}
\epsfig{file=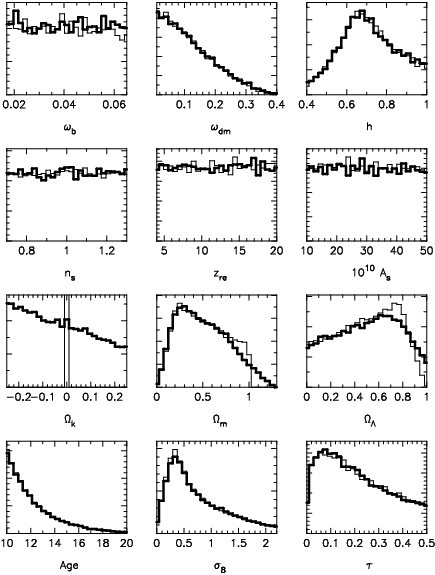,angle=-90,width=\linewidth}
\caption {The one-dimensional marginalised probability distributions for
cosmological parameters in Model A (thin line) and Model B (thick
line) using priors alone. Thinned samples from the MCMC chains are
plotted binned into histograms. The y axis shows the number of samples
in arbitrary units. }
\label{graph-implied}
\end{figure}

Given an ensemble of samples from the posterior probability
distributions, we can easily obtain one-dimensional marginalised
posterior distributions for any desired parameter, which thus
determine the constraints implied by the data.  Moreover, MCMC
techniques provide a natural way to perform model selection through
the evaluation of the Bayesian evidence (see e.g. Hobson, Bridle \&
Lahav 2002). For a given model $M$, the evidence is given by
\[
E \equiv \Pr(\vect{d}|M) 
= \int \Pr(\vect{d}|\btheta,M)\Pr(\btheta|M)\,d\btheta,
\]
which is simply the average of the likelihood over the prior.
The value of the evidence naturally incorporates
the spirit of Ockham's razor: a simpler theory, having a more
compact parameter space, will generally have a larger evidence than
a more complicated theory, unless the latter is 
significantly better at explaining the data. Thus, the problem of model
selection is answered simply by identifying the model with the largest
evidence. We have extended the \texttt{Cosmo-mc} program to include the
calculation of the evidence by thermodynamic integration (see e.g.
Hobson \& McLachlan 2002). This is implemented by numerically evaluating
the appropriate integral during the `burn-in' of the Markov
chain. 'Burn-in' samples are discarded for the purpose of the
parameter estimation.

The resulting software was run on a 24-node Linux cluster, with each
node propagating an independent chain. After burn-in, typically 5000
accepted samples were drawn from each chain, yielding a total of
120,000 accepted samples.  Since successive samples from a Markov chain
are, by nature, correlated, the accepted samples were thinned by a
factor of 15 before analysing them further; this resulted in 8,000
independent samples. We have found that our final results are very
insensitive to the thinning factor. Since each chain is run
independently, we also obtained 24 separate estimates of the evidence,
which are used to obtain an average value for $E$ and an estimate of
the associated error.

For calculation of confidence limits we use the $0.165$, $0.5$ and
$0.835$ points of the cumulative probability distribution. Thus, our
parameter estimate is the median of the
marginalised posterior pdf and the confidence interval 
encompasses 67\% of the probability. Using the median
instead of the maximum of the posterior has two advantages.
Firstly, it gives consistent results under monotonic parameter
transformation and combination (e.g. the median $\omega_b$ is equal to
the median $\Omega_b$ multiplied by the square of the median
$h$). This is especially advantageous in the case of cosmological
parameter estimation when it is not always clear which parameters
should be considered basic (for more information see
Jaynes 2002, p.621). Secondly, the method is very robust
when dealing with MCMC chains: the samples may be simply sorted and
searched with no need for binning and smoothing.  Of course, quoting
best estimates and confidence limits is simply a means of characterising
the full one-dimensional marginalised distributions for each parameter.
Finally, for Models A and B, the parameter estimation presented here
has been repeated independently using the standard grid-based method and the results 
are consistent. We note that the grid based approach would be
computationally prohibitive for more complex Models C and D.

\begin{figure}
\epsfig{file=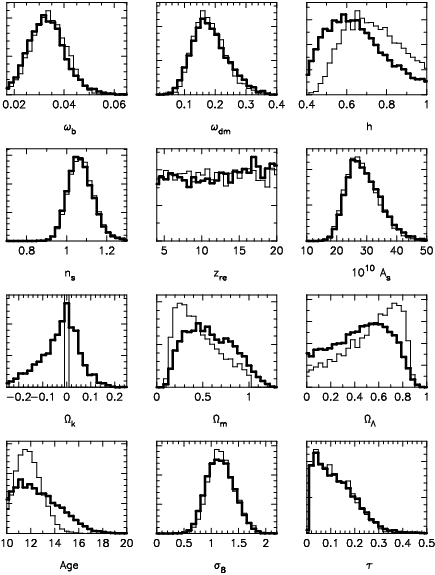,angle=-90,width=\linewidth}
\caption {Marginalised posterior probability distributions of
parameters from VSA and COBE data alone. The thin line corresponds to 
Model A and the thick line to Model B. Thinned samples from
the MCMC chains are plotted binned into histograms. The y axis shows the
number of samples in arbitrary units. The horizontal axes from
Fig.~\ref{prior} are retained. }
\label{vsae-only}
\end{figure}

\begin{figure}
\epsfig{file=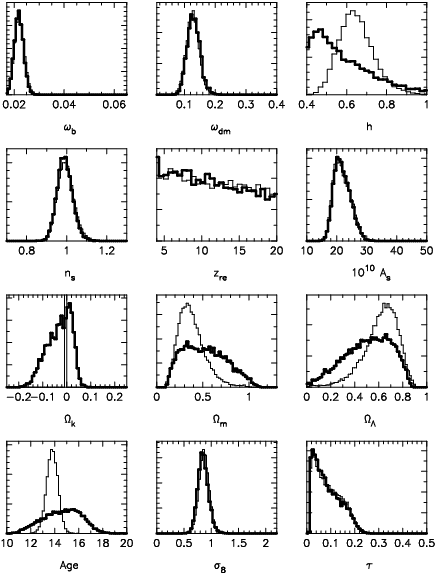,angle=-90,width=\linewidth}
\caption{As for Fig.~\ref{vsae-only} but using data from 
all considered CMB experiments.}
\label{cmb-only}
\end{figure}

\begin{figure}
\epsfig{file=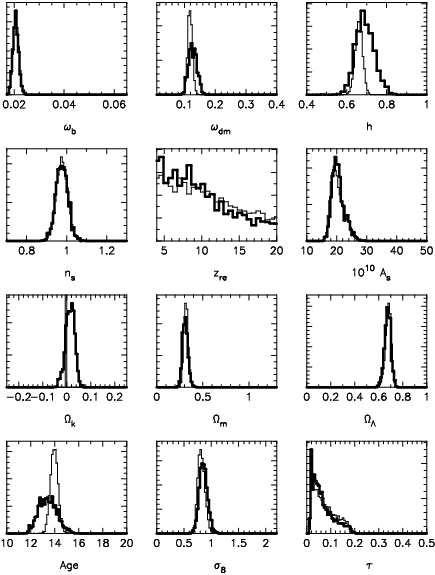,angle=-90,width=\linewidth}
\caption{As for Fig.~\ref{vsae-only} but using all the data sets
considered.}
\label{all-only}
\end{figure}

\subsection{Parameter constraints from priors alone}

The priors on the individual basic parameters are summarised in
Tables~\ref{prior} and \ref{prior2}. As noted earlier, we also adopt
the constraints (on derived parameters) $\Omega_\Lambda > 0$ and the
top-hat prior (10,20) Gyr on the age of the Universe. It is of
interest to determine the effect of these combined priors {\em alone}
on the one-dimensional marginalised posteriors for each parameter.
Using the MCMC sampler, this is easily performed by simply setting the
likelihood to a constant value, i.e. the
analysis is performed using no data. The resulting marginalised 
distributions for Model A and Model B are shown in Fig.~\ref{graph-implied}. 

These results are worthy of some discussion. Consider, for example,
the distribution for $h$. This has a clear peak around $h \approx
0.7$, which may be understood as follows.
Models with low $h$ are relatively disfavoured because they would
require $\Omega_\Lambda<0$ even for modest values of
$\omega_{\rm m}=\omega_{\rm dm}+\omega_{\rm b}$, 
while models with high $h$ are
disfavoured because they tend to have ages below 10 Gyr. 
Similar
arguments may be applied to explain the shapes of the other
distributions. In particular, we 
note that, although broad, the distributions for $\Omega_{\rm m}$ and
$\Omega_\Lambda$ peak at 0.3 and 0.7 respectively.
The non-uniformity of the
distributions in Fig.~\ref{graph-implied} represents the effect of the
sensible, and seemingly innocuous, initial constraints on the chosen
parameters. We have also repeated the analysis for Model D, and find
that the resulting marginalised distributions for the extra parameters 
$R$, $n_t$, $f_\nu$ and $w$ accurately follow the initial flat prior. 
Having made explicit the impact of our priors, we can now modulate
the above distributions by including increasing amounts of data from
various cosmological observations via the likelihood function.

\section{Results}

We consider three different sets of data in our analysis.  Firstly, we
use only VSA data (from both extended and compact configurations using
the main binning from Paper V), together with the COBE power spectrum
points (Smoot et al. 1992) to constrain the low-$\ell$ normalisation.
For this data set, we investigated just Models A and B.  Secondly, we
also include data from several recent CMB experiments, namely Maxima
(Hanany et al. 2002), Boomerang (Netterfield et al. 2002), DASI
(Halverson et al. 2002), CBI (Sievers et al. 2002) and Archeops
(Benoit et al. 2002).  For these data, we consider Models A, B and C.
Finally, we include a wide range of additional independent
cosmological probes to obtain the tightest possible constraints on the
model parameters. In addition to CMB observations, we include
constraints from: the HST Key Project (Freedman et al. 2001); the
first 147,000 redshifts measurements from the 2dF Survey (Colles et
al. 2001) on scales $0.02<k/(h
\mbox{Mpc}^{-1})<0.15$ (Lewis and Bridle 2002; Percival et al. 2002); nucleosynthesis (Burles,
Nollett \& Turner 2001); and type IA supernovae (Perlmutter et
al. 1999). We also include the constraint from the gas fraction in relaxed
clusters (Allen, Schmidt \& Fabian 2002; Allen et al. 2002). We do not
use the local X-ray luminosity function constraints from the cited
papers due to possible systematic uncertainties. Following
Allen et al., we marginalise over the cluster bias parameter.
Throughout the analysis we marginalise over applicable calibration and
beam uncertainty (see Paper IV).

\begin{figure}
\epsfig{file=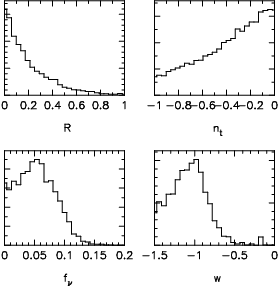,angle=-90,width=\linewidth}
\caption {Marginalised posterior probability distributions of
additional parameters for Model D using all data sets considered.}
\label{extrapars}
\end{figure}

\begin{figure}
\epsfig{file=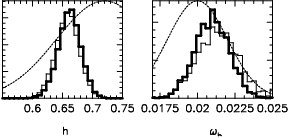,angle=-90,width=\linewidth}
\caption {Marginalised posterior probability distributions
for $h$ and $\omega_{\rm b}$ for Model A, using all cosmological
data, with (thick line) and without (thin line) the relevant 
direct constraints (dashed line) from the HST Key Project 
and nucleosynthesis respectively.}
\label{hbbn}
\end{figure}

The resulting one-dimensional marginalised posterior distributions,
for each of the three data sets, are shown in
Figs~\ref{vsae-only} -- \ref{all-only} 
for a selection of interesting parameters. 
The corresponding parameter estimates, confidence limits and evidences are 
given in Table \ref{results}. Note that the evidence values have been
scaled so that, for each data set, Model A has an evidence $E$ of unity.

\begin{table*}
\label{results}
\caption{Constraints on cosmological parameters; consult text for
details. If a parameter is constrained, the
confidence limits were obtained by calculating 0.165, 0.5 and 0.835
points of the cumulative probability distribution. When the marginalised posterior for a parameter 
does not contain a peak, 95\% confidence limits are given.
We also quote evidence values, but note that these
can only be compared between models when considering the same data set.}
\begin{tabular}{lccccc}
\hline\hline
&  Model A & Model B & Model A & Model B & Model C \\
          & (VSA \& COBE) & (VSA \& COBE) & (All CMB) & (All CMB) & (All CMB)\\
\hline
$\log E$ & $0.0 \pm 0.25$ & $-0.8 \pm 0.25$ & $0.0 \pm 0.03$ 
& $-1.3 \pm 0.03$ & $-3.1 \pm 0.6$ \\
\hline
$\omega_{b}$ &   $0.034\pm ^{ 0.007 }_{ 0.007}$   
&   $0.033\pm ^{ 0.007 }_{ 0.007}$   
&   $0.022\pm ^{ 0.002 }_{ 0.002}$   
&   $0.022\pm ^{ 0.002 }_{ 0.002}$   
&   $0.023\pm ^{ 0.002 }_{ 0.002}$  \\
$\omega_{dm}$ &   $0.18\pm ^{ 0.05 }_{ 0.04}$   
&   $0.18\pm ^{ 0.06 }_{ 0.05}$   &   $0.13\pm ^{ 0.02 }_{0.02}$   
&   $0.13\pm ^{ 0.02}_{ 0.02}$   &   $0.12\pm ^{ 0.02 }_{ 0.02}$  \\
$h$ &   $ 0.72\pm ^{  0.15 }_{  0.13}$   &   $ 0.63\pm ^{  0.16 }_{  0.13}$   
&   $ 0.64\pm ^{   0.09 }_{   0.07}$   &   $ 0.55\pm ^{  0.17 }_{  0.10}$   
&   $ 0.55\pm ^{  0.17 }_{  0.10}$  \\
$ n_{s}$ &   $1.07\pm ^{ 0.06 }_{ 0.06}$   &   $1.06\pm ^{ 0.06 }_{0.06}$   
&   $0.99\pm ^{ 0.04 }_{ 0.04}$   &   $0.99\pm ^{ 0.04 }_{ 0.03}$   
&   $1.03\pm ^{ 0.06 }_{ 0.05}$  \\
$z_{\rm re}$ &  unconstrained & unconstrained & unconstrained 
& unconstrained & unconstrained \\
$ 10^{10} A_{s}$ &   $28\pm ^{ 5 }_{ 4}$   
&   $28 \pm ^{ 6 }_{ 4}$   &   $21\pm ^{ 3 }_{ 2}$   
&   $22 \pm ^{ 3 }_{ 2}$   &   $21\pm ^{ 3 }_{ 2}$  \\
$R$ &      &      &      &      &   $0.22\pm ^{ 0.34 }_{ 0.17}$  \\
$ n_{t}$ &      &      &      &      &   $-0.29\pm ^{ 0.20 }_{ 0.35}$  \\
$\Omega_{k}$ &      &   $-0.02\pm ^{ 0.06 }_{ 0.09}$   &      
&   $-0.02\pm ^{ 0.04 }_{ 0.06}$   &   $-0.04\pm ^{ 0.05 }_{ 0.07}$  \\
$\Omega_{m}$ &   $0.42\pm ^{ 0.29 }_{ 0.18}$   
&   $0.56\pm ^{ 0.29 }_{ 0.24}$   &   $0.36\pm ^{ 0.15 }_{ 0.11}$   
&   $0.51\pm ^{ 0.26 }_{ 0.22}$   &   $0.46\pm ^{ 0.25 }_{ 0.20}$  \\
$ \Omega_{\Lambda}$ &   $0.58\pm ^{ 0.18 }_{ 0.29}$   
&   $0.47\pm ^{ 0.22 }_{ 0.28}$   &   $0.64\pm ^{ 0.11 }_{ 0.15}$   
&   $0.52\pm ^{ 0.18 }_{ 0.22}$   &   $0.59\pm ^{ 0.16 }_{ 0.21}$  \\
$\mbox{Age}$ &   $11.9\pm ^{ 1.1 }_{ 0.9}$   &   $12.5\pm ^{ 2.1 }_{1.5}$   
&   $13.8\pm ^{ 0.5 }_{ 0.4}$   &   $14.7\pm ^{ 1.4 }_{ 1.7}$   
&   $15.2\pm ^{ 1.6 }_{ 1.8}$  \\
$\sigma_{8}$ &   $1.17\pm ^{ 0.25 }_{ 0.20}$   
&   $1.16\pm ^{ 0.25 }_{ 0.23}$   &   $0.87\pm ^{ 0.09 }_{ 0.09}$   
&   $0.85\pm ^{ 0.10 }_{ 0.09}$   &   $0.81\pm ^{ 0.10 }_{ 0.10}$  \\
$\tau$ &   $ (0.01, 0.25) $   &   $ (0.01, 0.28) $   &   $ (0.02,0.19) $   
&   $ (0.01, 0.19) $   &   $ (0.02, 0.21) $  \\
\hline
&  & Model A & Model B & Model C  & Model D \\
&  & (All data) & (All data) & (All data) & (All data) \\
\hline
$\log E$ & & $0.0 \pm 0.4$ &  $-2.2 \pm 0.5$ & $-3.7 \pm 0.6$ &
$-6.7\pm 0.5$ \\
\hline
$\omega_{b}$ & &   $0.0210\pm ^{ 0.0011 }_{ 0.0011}$   
&   $0.0209\pm ^{ 0.0011 }_{ 0.0011}$   
&   $0.0215\pm ^{ 0.0012 }_{ 0.0011}$   
&   $0.0219\pm ^{ 0.0014 }_{ 0.0013}$  \\
$\omega_{dm}$ & &  $0.120\pm ^{ 0.008 }_{ 0.007}$   
&   $0.128\pm ^{ 0.014 }_{ 0.012}$   &   $0.119\pm ^{ 0.014 }_{0.014}$   
&   $0.128\pm ^{ 0.020 }_{ 0.020}$  \\
$h$ & &  $ 0.66\pm ^{   0.02 }_{  0.02}$   &   $ 0.69\pm ^{   0.05 }_{   0.04}$   
&   $ 0.68\pm ^{ 0.05 }_{ 0.05}$   &   $ 0.68\pm ^{   0.05 }_{   0.05}$  \\
$ n_{s}$ & &   $0.98\pm ^{ 0.03 }_{ 0.03}$   &   $0.98\pm ^{ 0.03 }_{0.03}$   
&   $1.00\pm ^{ 0.04 }_{ 0.03}$   &   $1.01\pm ^{ 0.05 }_{ 0.04}$  \\
$z_{re}$ & &  $ (4.00, 18.50) $   &   $ (4.00, 18.16) $   
&   $ (4.00, 17.84) $   &   $ (4.00, 17.95) $  \\
$ 10^{10} A_{s}$ & &   $20\pm ^{ 3 }_{ 2}$   
&   $20\pm ^{ 2 }_{ 2}$   &   $20\pm ^{ 2 }_{ 2 }$   
&   $20\pm ^{ 2 }_{ 2}$  \\
$R$ & &     &      &   $ (0.00, 0.53) $   &   $ (0.00, 0.63) $  \\
$ n_{t}$ & &      &      &   $ (-0.88, -0.00) $   &   $ (-0.88, -0.00) $  \\
$f_\nu$ &  &    &      &      &   $ (0.00, 0.11) $  \\
$ w$ &  &    &      &      &   $-1.06\pm ^{ 0.20 }_{ 0.25}$  \\
$\Omega_{k}$ & &      &   $0.02\pm ^{ 0.02 }_{ 0.02}$   
&   $0.00\pm ^{ 0.02 }_{ 0.02}$   &   $0.01\pm ^{ 0.03 }_{ 0.02}$  \\
$\Omega_{m}$ &  & $0.32\pm ^{ 0.02 }_{ 0.02}$   
&   $0.31\pm ^{ 0.03 }_{ 0.03}$   &   $0.30\pm ^{ 0.03 }_{ 0.03}$   &   $0.32\pm ^{ 0.06 }_{ 0.05}$  \\
$ \Omega_{\Lambda}$ & &   $0.68\pm ^{ 0.02 }_{ 0.02}$   
&   $0.67\pm ^{ 0.02 }_{ 0.02}$   &   $0.69\pm ^{ 0.03 }_{ 0.03}$   &   $0.66\pm ^{ 0.05 }_{ 0.06}$  \\
$\mbox{Age}$ &  & $14.0\pm ^{ 0.3 }_{ 0.3}$   
&   $13.4\pm ^{ 0.8 }_{ 0.7}$   &   $13.8\pm ^{ 0.9 }_{ 0.8}$   
&   $13.6\pm ^{ 1.0 }_{ 0.9}$  \\
$\sigma_{8}$ &  &  $0.82\pm ^{ 0.07 }_{ 0.06}$   
&   $0.86\pm ^{ 0.07 }_{ 0.07}$   &   $0.83\pm ^{ 0.08 }_{ 0.08}$   
&   $0.71\pm ^{ 0.09 }_{ 0.09}$  \\
$\tau$ &  & $ (0.02, 0.17) $   &   $ (0.02, 0.16) $   &   $ (0.02,0.16) $   
&   $ (0.01, 0.16) $  \\
\hline\hline
\end{tabular}
\end{table*}

\section{Discussion and Conclusions}

By comparing the effective priors on the cosmological parameters
plotted in Fig.~\ref{graph-implied} with the posterior distributions
inferred from the VSA and COBE data alone (Fig.~\ref{vsae-only}), we
see that the main strength of the VSA lies in its large $\ell$-range,
which allows one to constrain a wide variety of cosmological
parameters. In particular, we observe that the VSA data significantly
improves the constraints on $\omega_{\rm b}$, $\omega_{\rm dm}$,
$\Omega_{\rm k}$, $\sigma_8$ and $n_s$.  We note that, as a direct
consequence of the pronounced third peak in the VSA power spectrum,
the preferred value for $\omega_{\rm b}$ is somewhat larger than that
from the nucleosynthesis constraint (Burles et al. 2001) or that from
the combined CMB data. This excess is statistically significant only
at the $1.6 \sigma$ level. VSA and COBE data alone do not, however,
provide significant new constraints on either $h$ or $\Omega_\Lambda$,
especially for Model B.

On comparing Figs~\ref{vsae-only} and \ref{cmb-only}, we see
that, as one would expect, the constraints on all parameters are
tightened by the inclusion of all the CMB data. 
Indeed, for Model A, reasonable constraints are obtained on
all the parameters under consideration, except for $z_{\rm re}$,
although there is some indication that low values of $z_{\rm re}$ are
preferred. For many of the parameters, the constraints are not
significantly broader when one allows the spatial curvature to vary in
Model B. Nevertheless, for this model, some parameters do become 
relatively unconstrained, in particular $h$, $\Omega_{\rm m}$ and
$\Omega_\Lambda$, and the limits on the age of the Universe are
widened considerably.

Using all recent available cosmological 
data sets allows one to place tight constraints on nearly 
all parameters, even for Model B. Indeed, only $z_{\rm re}$ remains
relatively unconstrained, but the earlier indication that lower values
are preferred appears to be reinforced. We note that, aside from the
parameters $h$ and the age of the Universe, the constraints obtained
are very similar for Model A and Model B. In fact, from
Table~\ref{results}, we see that, even for Models C and D, the
constraints on the parameters plotted in Fig.~\ref{all-only} are not 
significantly broadened. In Fig.~\ref{extrapars}, we also plot
the constraints obtained on the additional parameters
$R$, $n_t$, $f_\nu$ and $w$ for Model D. We see that the data favour
low values of $R$ and high values of $n_t$. More significant
constraints can be placed on $f_\nu$ and $w$. We note from the figure
that the marginalised distribution for each parameter possesses a
single peak, although these are not pronounced features, especially
for $f_\nu$. Of particular interest is that the preferred value of $w$
corresponds to a cosmological constant and that
a firm upper limit is obtained on the fraction of dark matter 
in the form of massive neutrinos. 

To distinguish between the different cosmological models considered
here, we have calculated the Bayesian evidence for each model-data set
combination. It must be remembered that evidence values can only be
compared between different models using the same data set.  From
Table~\ref{results}, we see that, for each data set, Model A (i.e. the
simple 6-parameter $\Lambda$CDM cosmology) is preferred, although the
evidence ratio between Model A and Model B is of order unity in each
case. For Models C and D, however, the evidence ratio with Model A is
considerably larger and shows clearly that such general models are not
necessary to explain current cosmological observations.

Since Model A is found to have the largest evidence, it is of interest
to consider more closely the impressively tight parameter constraints
that can be achieved in this simple case using all the available
cosmological data. In particular, let us focus on $h$ and
$\omega_{\rm b}$, which are parameters directly probed by the HST
Key Project (Freedman et al. 2001) and nucleosynthesis data (Burles et
al. 2001). In Fig.~\ref{hbbn}, we plot the posterior distributions
of these parameters for Model A, with (thick line) and without (thin
line) the corresponding direct constraint (dashed line). In each case,
we see that the direct constraint is overwhelmed by the combination
of the other data sets.

It is clear from the above analyses that the
constraints on the parameters in the simplest models
become impressively small when a wide range of cosmological probes
are used. It remains to be seen, however, whether this 
level of precision is matched by a corresponding level of 
accuracy in the values determined. The model comparison
exercise demonstrated here does not address the possibility
that different experiments have different systematic
errors. Nevertheless, one would hope that these might be reduced when 
many experiments are combined. In future work, we will investigate
other models, with different combinations of free and fixed parameters
and perform a hyperparameter analysis (Hobson et
al. 2002) to reveal any discrepancies between data sets.

\section*{ACKNOWLEDGEMENTS} 

We thank Antony Lewis and Sarah Bridle for providing their
\texttt{Cosmo-mc} code and many useful discussions regarding MCMC
parameter estimation. We thank Carolina \"Odman for useful
discussions. We also thank the staff of the Mullard Radio Astronomy
Observatory, the Jodrell Bank Observatory and the Teide Observatory
for invaluable assistance in the commissioning and operation of the
VSA. The VSA is supported by PPARC and the IAC. RS, YH, KL, and PM
acknowledge support by PPARC studentships. KC acknowledges a Marie
Curie Fellowship. YH is supported by the Space Research Institute of
KACST.  AS acknowledges the support of St. Johns College, Cambridge.
We thank Professor Jasper Wall for assistance and advice throughout
the project. 


\bsp
\label{lastpage}
\end{document}